\documentclass[runningheads]{llncs}
\usepackage[T1]{fontenc}
\usepackage[dvipsnames,table]{xcolor}
\usepackage{graphicx}
\usepackage[hyphens]{url}
\usepackage[hidelinks,draft=false]{hyperref}
\usepackage{amsmath,amssymb}
\usepackage{amsfonts}
\usepackage{tikz}
\usetikzlibrary{positioning,shadows,fit,calc,shapes}
\usepackage{pgf-umlsd}
\usepackage[inline]{enumitem}
\setlist{noitemsep}
\usepackage{breakcites}
\usepackage{soul}
\usepackage{listings}
\usepackage{array,etoolbox}
\usepackage{multirow}
\usepackage{algorithm}
\usepackage[noend]{algpseudocode}
\usepackage{fancyvrb}
\usepackage{xcolor}
\preto\tabular{\setcounter{magicrownumbers}{0}}
\newcounter{magicrownumbers}
\newcommand\rownumber{\stepcounter{magicrownumbers}\arabic{magicrownumbers}}
\newcommand{\alert}[1]{#1}

\newcommand{\cmmnt}[1]{\ignorespaces} 
\newcommand{\kwl}[1]{\mathbf{#1}}
\newcommand{\kwf}[1]{\mathsf{#1}}
\newcommand{\kwc}[1]{\mathsf{#1}}

\newcommand{\kwt}[1]{\mathsf{#1}}
\newcommand{\kwe}[1]{\mathsf{#1}}

\newcommand{\var}[1]{\mathit{#1}}
\newcommand{\param}[1]{\texttt{#1}}
\newcommand{\ok}{$\checkmark$}
\newcommand{\no}{$\times$}

\newcommand{\tca}[1]{\textcolor{purple}{#1}}
\newcommand{\tcb}[1]{\textcolor{orange}{#1}}
\newcommand{\tcc}[1]{\textcolor{teal}{#1}}

\newcommand{\circlednumber}[1]{{#1.}}
\newcommand*\circled[1]{\tikz[baseline=(char.base)]{
    \node[shape=ellipse,draw,inner sep=0.6pt] (char) {#1};}}

\makeatletter
\renewcommand{\ALG@beginalgorithmic}{\scriptsize}

\makeatother
\makeatletter
\def\blfootnote{\xdef\@thefnmark{}\@footnotetext}
\makeatother
\begin{document}

\title{Bulwark: Holistic and Verified Security Monitoring of Web Protocols}

\titlerunning{Holistic and Verified Security Monitoring of Web Protocols}

\author{Lorenzo Veronese\inst{1,2}\fnmsep\thanks{Now at TU Wien. Corresponding author: \texttt{lorenzo.veronese@tuwien.ac.at}.} 
   \and Stefano Calzavara\inst{1} 
   \and Luca Compagna\inst{2}}

\authorrunning{Veronese, Compagna, Calzavara}

\institute{Università Ca' Foscari Venezia \and SAP Labs France}

\maketitle              
\begin{abstract}
Modern web applications often rely on third-party services to provide their functionality to users. The secure integration of these services is a non-trivial task, as shown by the large number of attacks against Single Sign On and Cashier-as-a-Service protocols. In this paper we present Bulwark, a new automatic tool which generates formally verified security monitors from applied pi-calculus specifications of web protocols. The security monitors generated by Bulwark offer holistic protection, since they can be readily deployed both at the client side and at the server side, thus ensuring full visibility of the attack surface against web protocols. We evaluate the effectiveness of Bulwark by testing it against a pool of vulnerable web applications that use the OAuth 2.0 protocol or integrate the PayPal payment system.

\end{abstract}
%
%
\begingroup
\let\clearpage\relax

\section{Introduction}\label{sec:introduction}

Modern web applications often rely on third-party services to provide their functionality to users. The trend of integrating an increasing number of these services has turned traditional web applications into \emph{multi-party} web apps (MPWAs, for short) with at least three communicating actors. In a typical MPWA, a \textit{Relying Party} (RP) integrates services provided by a \textit{Trusted Third Party} (TTP). Users interact with the RP and the TTP through a \textit{User Agent} (UA), which is normally a standard web browser executing a \emph{web protocol}. For example, many RPs authenticate users through the Single Sign On (SSO) protocols offered by TTPs like Facebook, Google or Twitter, and use Cashier-as-a-Service (CaaS) protocols provided by payment gateway services such as PayPal and Stripe. 

Unfortunately, previous research showed that the secure integration of third-party services is a non-trivial task~\cite{wang2012,wang2011,sun2012,fett2016,webspiCSF,Pellegrino2014,BLAST}. Vulnerabilities might arise due to errors in the protocol specification~\cite{fett2016,webspiCSF}, incorrect implementation practices at the RP~\cite{wang2011,sun2012,Pellegrino2014} and subtle bugs in the integration APIs provided by the TTP~\cite{wang2012}. To secure MPWAs, researchers proposed different approaches, most notably based on \emph{runtime monitoring}~\cite{WPSE,oauthguard,integuard,aegis,block}. The key idea of these proposals is to automatically generate security monitors allowing only the web protocol runs which comply with the expected, ideal run. Security monitors can block or try to automatically fix protocol runs which deviate from the expected outcome.

In this paper, we take a retrospective look at the design of previous proposals for the security monitoring of web protocols and identify important limitations in the current state of the art. In particular, we observe that:
\begin{enumerate}
    \item existing proposals make strong assumptions about the placement of security monitors, by requiring them to be deployed either at the client~\cite{WPSE,oauthguard} or at the RP~\cite{integuard,aegis,block}. In our work we show that both choices are sub-optimal, as they cannot prevent all the vulnerabilities identified so far in popular web protocols (see Section~\ref{sec:designspace});

    \item most existing proposals are not designed with formal verification in mind. They can ensure that web protocol runs are compliant with the expected run, e.g., derived from the network traces collected in an unattacked setting, however they do not provide any guarantee about the actual security properties supported by the expected run itself~\cite{integuard,aegis,block}.
\end{enumerate}

Based on these observations, we claim that none of the existing solutions provides a reliable and comprehensive framework for the security monitoring of web protocols in MPWAs.

\paragraph*{Contributions.}
In this paper, we contribute as follows:
\begin{enumerate}
    \item we perform a systematic, comprehensive design space analysis of previous work and we identify concrete shortcomings in all existing proposals by considering the popular OAuth 2.0 protocol and the PayPal payment system as running examples (Section~\ref{sec:designspace});
    
    \item we present Bulwark, a novel proposal exploring a different point of the design space. Bulwark generates formally verified security monitors from applied pi-calculus specifications of web protocols and lends itself to the appropriate placement of such monitors to have full visibility of the attack surface, while using modern web technologies to support an easy deployment. This way, Bulwark reconciles formal verification with practical security (Section~\ref{sec:approach});
    
    \item we evaluate the effectiveness of Bulwark by testing it against a pool of vulnerable web applications that use the OAuth 2.0 protocol or integrate the PayPal payment system. Our analysis shows that Bulwark is able to successfully mitigate attacks on both the client and the server side (Section~\ref{sec:evaluation}).
    
\end{enumerate}

\section{Motivating Example}\label{sec:example}
As motivating example, shown in \figurename~\ref{fig:oauth-example}, we selected a widely used web protocol, namely OAuth 2.0 in explicit mode, which allows a RP to leverage a TTP for authenticating a user operating a UA.\footnote{The protocol in the figure closely follows the Facebook implementation; details might slightly vary for different TTPs.}
The protocol starts (step 1) with the UA visiting the RP's login page. A login button is provided back that, when clicked, triggers a request to the TTP (steps 2-3). 
Such a request comprises: \param{client\_id}, the identifier registered for the RP at the TTP;  \param{reduri}, the URI at RP to which the TTP will redirect the UA after access has been granted; and \param{state}, a freshly generated value used by the RP to maintain session binding with the UA. The UA authenticates with the TTP (steps 4-5), which in turn redirects the UA to the \param{reduri} at RP with a freshly generated value \param{code} and the \param{state} value (steps 6-7). 
The RP verifies the validity of \param{code} in a back channel exchange with the TTP (steps 8-9): the TTP acknowledges the validity of \param{code} by sending back a freshly generated \param{token} indicating the UA has been authenticated. Finally, the RP confirms the successful authentication to the UA (step 10).

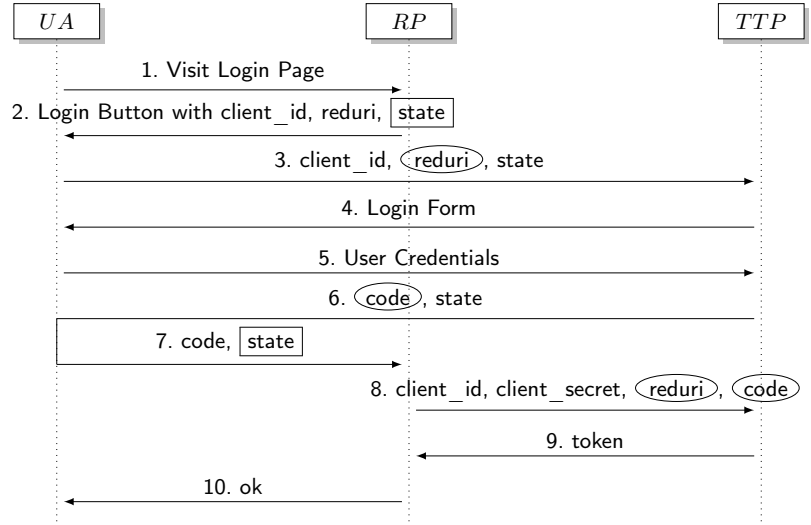
\begin{figure}[!t]
\centering
\resizebox{0.9\textwidth}{!}{
\def\hd{0.65cm}
\def\sp{0.1cm}
\begin{tikzpicture}[
    node distance=5cm and 1cm,
    font=\sffamily\footnotesize,
    box/.style={rectangle, draw ,text width = 1cm, fill=white, drop shadow,
      align=center, minimum height = 0.5cm},
    mess/.style={font=\sffamily\scriptsize}
    ]
    \node [box] (ua) {$UA$};
    \node [box, right of = ua] (rp) {$RP$};
    \node [box, right of = rp] (ttp) {$TTP$};
    \path[draw, -latex] ($(ua)-(-\sp,1.5*\hd)$) to node[above]{\circlednumber{1} Visit Login Page} ($(rp)-(\sp,1.5*\hd)$);
    \path[draw, latex-] ($(ua)-(-\sp,2.5*\hd)$) to node[above]{\circlednumber{2} Login Button with client\_id, reduri, \fbox{state}} ($(rp)-(\sp,2.5*\hd)$);
    \path[draw, -latex] ($(ua)-(-\sp,3.5*\hd)$) to node[above]{\circlednumber{3} client\_id, \circled{reduri}, state} ($(ttp)-(\sp,3.5*\hd)$);
    \path[draw, latex-] ($(ua)-(-\sp,4.5*\hd)$) to node[above]{\circlednumber{4} Login Form} ($(ttp)-(\sp,4.5*\hd)$);
    \path[draw, -latex] ($(ua)-(-\sp,5.5*\hd)$) to node[above]{\circlednumber{5} User Credentials} ($(ttp)-(\sp,5.5*\hd)$);
    \path[draw, -] ($(ua)-(0,6.5*\hd)$) to node[above]{\circlednumber{6} \circled{code}, state} ($(ttp)-(\sp,6.5*\hd)$);
    \path[draw, -] ($(ua)-(0,6.5*\hd)$) to node[above]{} ($(ua)-(0,7.5*\hd)$);
    \path[draw, -latex] ($(ua)-(0,7.5*\hd)$) to node[above]{\circlednumber{7} code, \fbox{state}} ($(rp)-(\sp,7.5*\hd)$);
    \path[draw, -latex] ($(rp)-(-\sp,8.5*\hd)$) to node[above]{\circlednumber{8} client\_id, client\_secret, \circled{reduri}, \circled{code} } ($(ttp)-(\sp,8.5*\hd)$);
    \path[draw, latex-] ($(rp)-(-\sp,9.5*\hd)$) to node[above]{\circlednumber{9} token } ($(ttp)-(\sp,9.5*\hd)$);
    \path[draw, latex-] ($(ua)-(-\sp,10.5*\hd)$) to node[above]{\circlednumber{10} ok } ($(rp)-(\sp,10.5*\hd)$);
    \path[draw, dotted] (ua) -- ($(ua)-(0,11*\hd)$);
    \path[draw, dotted] (rp) -- ($(rp)-(0,11*\hd)$);
    \path[draw, dotted] (ttp) -- ($(ttp)-(0,11*\hd)$);
\end{tikzpicture}}
\caption{Motivating example: Facebook OAuth 2.0 explicit mode}
\label{fig:oauth-example}
\end{figure}

Securely implementing such a protocol is far from easy and many vulnerabilities have been reported in the past. We discuss below two representative attacks with severe security implications.

\subsubsection{Session Swapping~\cite{sun2012}.}
\label{sec:atk1}
Session swapping exploits the lack of contextual binding between the login 
endpoint (step 2) and the callback endpoint (step 7). This is often the case in RPs that do
not provide a \param{state} parameter or do not strictly validate it.
The attack starts with the attacker signing in to the TTP and obtaining a valid
\param{code} (step 6). The attacker then tricks an honest user, through CSRF, to send the
attacker's \param{code} to the RP, which makes the victim's UA authenticate at the RP with the attacker's identity. From there on, the attacker can track the activity of the victim at the RP. The RP can prevent this attack by checking that the value of \param{state} at step 7 matches the one that was generated at step 2.  The boxed shapes around \param{state} represent this invariant in \figurename~\ref{fig:oauth-example}.

\subsubsection{Unauthorized Login by Code Redirection~\cite{webspiCSF,RFC6749}.} 
\label{sec:atk2}
Code (and token) redirection attacks exploit the lack of strict validation of the \param{reduri}
parameter and involve its manipulation by the attacker. The attacker crafts a malicious page which fools the victim into starting the protocol flow at step 3 with valid \param{client\_id} and \param{state} from an honest RP, but with a \param{reduri} that points to the attacker's site. The victim then authenticates at the TTP and is redirected to the attacker's site with the \param{code} value. The attacker can then craft the request at step 7 with the victim's \param{code} to obtain the victim's \param{token} (step 9) and authenticate as her at the honest RP. The TTP can prevent this attack by (i) binding the \param{code} generated at step 6 to the \param{reduri} received at step 3, and (ii) checking, at step 8, that the received \param{code} is correctly bound to the supplied \param{reduri}. The rounded shapes represent this invariant in \figurename~\ref{fig:oauth-example}.

\section{Related Work}\label{sec:relatedwork}
We review here existing approaches to the security monitoring of web protocols, focusing in particular on their adoption in modern MPWAs. Each approach can be classified based on the placement of the proposed defense, i.e., we discriminate between client-side and server-side approaches. 

We highlight here that none of the proposed approaches is able to protect the entire attack surface of web protocols. Moreover, none of the proposed solutions, with the notable exception of WPSE~\cite{WPSE}, is designed with formal verification in mind and provides clear, precise guarantees about the actual security properties satisfied by the enforced policy.

\subsection{Client-Side Defenses}
WPSE~\cite{WPSE} extends the browser with a security monitor for web protocols
that enforces the intended protocol flow, as well as the confidentiality and the integrity of messages. 
This monitor is able to mitigate many vulnerabilities found in the literature.
The authors, however, acknowledge that some classes of attack cannot be prevented by WPSE. In particular, network attacks (like the HTTP variant of the IdP mix-up attack~\cite{fett2016}), attacks that do not deviate from the intended protocol flow (like the automatic login CSRF from~\cite{webspiCSF}) and purely server-side attacks are out of scope.

OAuthGuard~\cite{oauthguard} is a browser extension that aims to prevent five types of attacks on OAuth 2.0 and OpenID Connect, including CSRF and impersonation attacks. 
OAuthGuard essentially shares the same limitations of WPSE, due to the same partial visibility of the attack surface (the client side). 

Recently Google has shown interest in extending its Chrome browser to monitor SSO protocols,\footnote{\url{https://gsuiteupdates.googleblog.com/2018/04/more-secure-sign-in-chrome.html}}
however their solution deals with a specific attack against their own implementation of SAML and is not a general approach designed to protect other protocols or TTPs.

\subsection{Server-Side Defenses}
InteGuard~\cite{integuard} focuses on the server side of the RP, as its code appears to be more error-prone than that of the TTP. InteGuard is deployed as a proxy in front of the RP that checks invariants within the HTTP messages before they reach the web server. Different types of invariants are automatically generated from the network traces of SSO and CaaS protocols and enable the monitor to link together multiple HTTP sessions in transactions. Thanks to its placement, InteGuard can also monitor back channels (cf. steps 8-9 of \figurename~\ref{fig:oauth-example}).
The authors explicitly mention that InteGuard does not offer protection on the TTP, expecting further efforts to be made in that direction. Unfortunately, several attacks can only be detected at the TTP, e.g., some variants of the unauthorized login by auth. code (or token) redirection attack from~\cite{webspiCSF}. 

AEGIS~\cite{aegis} synthesizes runtime monitors to enforce control-flow and data-flow integrity, authorization policies and constraints in web applications. The monitors are server-side proxies generated by extracting invariants from a set of input traces. AEGIS was designed for traditional two-party web applications, hence it does not offer comprehensive protection in the multi-party setting, e.g., due to its inability to monitor messages exchanged on the back channels. However, as mentioned by the authors, it can still mitigate those vulnerabilities which can be detected on the front channel of the RP (e.g., the shop-for-free TomatoCart example~\cite{aegis}). Similar considerations apply to BLOCK~\cite{block}, a black-box approach for detecting \textit{state violation} attacks, i.e., attacks which exploit logic flaws in the application to allow some functionality to be accessed at inappropriate states.

Guha et al.~\cite{Guha2009} apply a static control-flow analysis for JavaScript code to construct a \textit{request-graph}, a model of a well-behaved client as seen by the server application. They then use this model to build a reverse proxy that blocks the requests that violate the expected
control flow of the application, and are thus marked as potential attacks. Also this approach was designed for two-party web applications, hence does not offer holistic protection in the multi-party setting. Moreover, protection can only be enforced on web applications which are entirely developed in JavaScript.

\section{Design Space Analysis}\label{sec:designspace}
Starting from our analysis of related work, we analyze the design space of security monitors for web protocols, discussing pros and cons along different axes. Our take-away message is that solutions which assume a fixed placement of a single security monitor, which is the path taken by previous work, are inherently limited in their design for several reasons. 

\subsection{Methodology}
We consider three possible deployment options for security monitors: the first two are taken from the literature, while the last one is a novel proposal we make. In particular, we focus on:
\begin{enumerate}
    \item \emph{browser extensions}~\cite{WPSE,oauthguard}: a browser extension is a plugin module, typically written in JavaScript, that extends the web browser with custom code with powerful capabilities on the browser internals, e.g., arbitrarily accessing the cookie jar and monitoring network traffic;
    \item \emph{server-side proxies}~\cite{integuard,aegis,block}: a proxy server acts as an intermediary sitting between the web server hosting (part of) the web application and the clients that want to access it; 
    \item \emph{service workers}: the Service Worker API\footnote{\url{https://developer.mozilla.org/en-US/docs/Web/API/Service_Worker_API}} is a new browser functionality that enables websites to define JavaScript workers running in the background, decoupled from the web application logic. Service workers provide useful features normally not available to JavaScript, e.g., inspecting HTTP requests before they leave the browser, hence are an intriguing deployment choice for client-side security monitors.
\end{enumerate}

We evaluate these options with respect to four axes, originally proposed as effective criteria for the analysis of web security solutions~\cite{CalzavaraFST17}:
\begin{enumerate}
    \item \emph{ease of deployment}: the practicality of a large-scale deployment of the defense mechanism, i.e., the price to pay for site operators to grant security benefits;
    \item \emph{usability}: the impact on the end-user experience, e.g., whether the user is forced to take actions to activate the protection mechanism;
    \item \emph{protection}: the effectiveness of the defense mechanism, i.e., the supported and unsupported security policies;
    \item \emph{compatibility}: the precision of the defense mechanism, i.e., potential false positives and breakages coming from security enforcement.
\end{enumerate}

\subsection{Ease of Deployment and Usability}\label{sec:easeofdeployment}
Service workers are appealing, since they score best with respect to ease of deployment and usability. Specifically, the deployment of a service worker requires site operators to just add a JavaScript file to the web application: when a user visits the web application, the installation of the service worker is transparently performed with no user interaction.

Server-side proxies similarly have the advantage of ensuring transparent protection to end users. However, they are harder to deploy than service workers, as they require site operators to have control over the server networking. Even if site operators just needed to apply small modifications to the monitored application, they would have to reroute the inbound/outbound traffic to the proxy. This is typically easy for the TTP, which is usually a major company with full control over its deployment, but it can be impossible for some RPs. RPs are sometimes 
deployed on managed hosting platforms that may not allow any modification on the server itself, except for the application code. 
Note that site operators could implement the logic of the proxy directly in the application code, but this solution is impractical, since it would require a significant rewriting of the web application logic. This is also particularly complicated when the web application is built on top of multiple programming languages and frameworks.

Finally, browser extensions are certainly the worst solution with respect to usability and ease of deployment. Though installing a browser extension is straightforward, site operators cannot assume that every user will perform this manual installation step. In principle, site operators could require the installation of the browser extension to access their web application, but this would have a major impact on usability and could drive users away. Users' trust in browser extensions is another problem on its own: extensions, once installed and granted permissions, have very powerful capabilities on the browser internals. Moreover, the extension should be developed for the plethora of popular browsers which are used nowadays. Though major browsers now share the same extension architecture, many implementation details are different and would force developers to release multiple versions of their extensions, which complicates deployment. In the end, installing an extension is feasible for a single user, but relying on browser extensions for a large-scale security enforcement is unrealistic.

\subsection{Protection and Compatibility}
We study protection and compatibility together, since the \emph{visibility} of the attack surface is the key enabler of both protection and compatibility. Indeed, the more the monitor has visibility of the protocol messages, the more it becomes able to avoid both false positives and false negatives in detecting potential attacks. 

We use the notation $P_1 \leftrightarrow P_2$ to indicate the channel between two parties $P_1$ and $P_2$. If a monitor has visibility over a channel, then the monitor has visibility over all the messages exchanged on that channel.   

\subsubsection{Visibility.} 
Browser extensions run on the UA and can have visibility of all the messages channeled through it. In particular, browser extensions can request \emph{host permissions} upon installation to get access to the traffic exchanged with arbitrary hosts, which potentially enables them to inspect and edit any HTTP request and response relayed through the UA. In MPWAs, the UA can thus have visibility over both the channels UA $\leftrightarrow$ RP and UA $\leftrightarrow$ TTP (shortly indicated as UA $\leftrightarrow$ \{RP, TTP\}). However, the UA itself is not in the position to observe the entire attack surface against web protocols: for example, when messages are sent on the back channel between the RP and the TTP (RP $\leftrightarrow$ TTP) like in our motivating example (steps 8-9 in \figurename~\ref{fig:oauth-example}), an extension is unable to provide any protection, as the UA is not involved in the communication at all.

Server-side proxies can be categorized into \emph{reverse proxies} and \emph{forward proxies}, depending on whether they monitor incoming or outgoing HTTP requests respectively (plus their corresponding HTTP responses). Both approaches are useful and have been proposed in the literature, e.g., InteGuard~\cite{integuard} uses both a reverse proxy and a forward proxy at the RP to capture messages from the UA and to the TTP respectively. This way, InteGuard has full visibility of all the messages flowing through the RP, i.e., RP $\leftrightarrow$ \{UA, TTP\}. However, this is still not sufficient to fully monitor the attack surface. In particular, server-side proxies cannot inspect values that never leave the UA, like the \emph{fragment identifier}, which instead plays an important role in the implicit flow of OAuth 2.0.

Finally, web applications can register service workers at the UA by means of JavaScript. Service workers can act as network proxies with access to all the traffic exchanged between the UA and the origin\footnote{An origin is a triple including a scheme (HTTP, HTTPS, ...), a host (\url{www.foo.com}) and a port (80, 443, ...). Origins represent the standard web security boundary.} which registered them. This way, service workers have the same visibility of a reverse-proxy sitting at the server; however, since they run on the UA, they also have access to values which never leave the client, like the fragment identifier. Despite this distinctive advantage, service workers are severely limited by the Same Origin Policy (SOP). In particular, they cannot monitor traffic exchanged between the UA and other origins, which makes them less powerful than browser extensions. For example, contrary to browser extensions like WPSE~\cite{WPSE}, a single service worker cannot monitor and defend both the RP and the TTP. This limitation can be mitigated by using multiple service workers and/or selectively relaxing SOP using CORS, which however requires collaboration between the RP and the TTP. Since service workers are more limited than browser extensions, they also share their inability to monitor back channels, hence they cannot be a substitute for forward proxies.

In the end, we conclude that browser extensions and server-side proxies are \emph{complementary} in their ability to observe security-relevant protocol components, given their respective positioning, while service workers are strictly less powerful than their alternatives.

\subsubsection{Cross-Site Scripting (XSS).}
XSS is a dangerous vulnerability which allows an attacker to inject malicious JavaScript code in a benign web application. Once a web application suffers from XSS, most confidentiality and integrity guarantees are lost, hence claiming security despite XSS is wishful thinking. Nevertheless, we discuss here how browser extensions can offer better mitigation than service workers in presence of XSS vulnerabilities. Specifically, since service workers can be removed by JavaScript, an attacker who was able to exploit an XSS vulnerability would also be able to void the protection offered by service workers. This can be mitigated by defensive programming practices, e.g., overriding the functions required for removing service workers, but it is difficult to assess both the correctness and the compatibility impact of this approach. For example, the deactivation of service workers might be part of the legitimate functionality of the web application or the XSS could be exploited before security-sensitive functions are overridden. Browser extensions, instead, cannot be removed by JavaScript and are potentially more robust against XSS. For example, WPSE~\cite{WPSE} replaces secret values with random placeholders before they actually enter the DOM, so that secrets exchanged in the monitored protocol cannot be exfiltrated via XSS; the placeholders are then replaced with the intended values only when they leave the browser towards authorized parties.

\subsubsection{Tamper Resistance.}
Since both browser extensions and service workers are installed on the client, they can be tampered with or uninstalled by malicious users or software. This means that the defensive practices put in place by browser extensions and service workers are voided when the client cannot be trusted. This is particularly important for applications like CaaS, where malicious users might be willing to abuse the payment system, e.g., to shop for free. Conversely, server-side proxies are resilient by design to this kind of attacks, since they cannot be accessed at the client side, hence they are more appropriate for web applications where the client cannot be trusted to any extent.

\subsubsection{Assessment on MPWAs.}
\label{sec:attacks}
We now substantiate our general claims by means of a list of known attacks against the OAuth 2.0 protocol and the PayPal payment system, two popular protocols in MPWAs. Table~\ref{tab:attacks} shows this list of attacks. For each attack, we show which channels need to be visible to detect the attack and we conclude whether the attack can be prevented by a browser extension (\textit{ext}), a service worker (\textit{sw}) or a server-side proxy (\textit{proxy}) deployed on either the RP or the TTP. 

\begin{table}[t!]
\newcommand{\indirect}{$\checkmark$}
\begin{center}
\resizebox{\textwidth}{!}{
\begin{small}
\begin{tabular}{@{\makebox[1.6em][r]{\rownumber\space}}|l|c|c|cc|cc}
\multicolumn{1}{@{} c| }{
\textbf{Attack}} & \textbf{Channels} &\textbf{UA} & \multicolumn{2}{c|}{\textbf{RP}} & \multicolumn{2}{c}{\textbf{TTP}} \\
\multicolumn{1}{@{} l| }{}
& \textbf{to observe} & \emph{ext} & \emph{sw} & \emph{proxy} & \emph{sw} & \emph{proxy}\\
\hline
\multicolumn{7}{c}{\textit{OAuth 2.0}}\\
\hline
307 Redirect attack \cite{fett2016}\cmmnt{(3.1\cite{fett2016})}
& UA$\leftrightarrow$TTP & \ok & \no & \no & \ok & \ok\\
Access token eavesdropping \cite{sun2012}\cmmnt{(A1\cite{sun2012})}
& UA$\leftrightarrow$RP \cmmnt{\alert{UA$\leftrightarrow$\{RP,TTP\}}} \cmmnt{UA$\leftrightarrow$RP, UA$\leftrightarrow$TTP} & \ok & \ok & \ok & \no & \no\\
Code/State Leakage via referer header \cite{fett2017,fett2016}\cmmnt{(III\cite{fett2017}) (3.3\cite{fett2016})}
& UA$\leftrightarrow$RP  & \ok & \ok & \ok & \no & \no\\
Code/Token theft via XSS \cite{sun2012}\cmmnt{(A2\cite{sun2012})} 
& UA$\leftrightarrow$RP & \ok & \no & \no & \no & \no\\
Cross Social-Network Request Forgery \cite{webspiCSF} \cmmnt{(V.D\cite{webspiCSF})}
& UA$\leftrightarrow$RP & \ok & \ok & \ok & \no & \no\\
Facebook implicit AppId Spoofing \cite{wang2012,BLAST} \cmmnt{(4.2\cite{wang2012}, \#6\cite{BLAST})} 
& UA$\leftrightarrow$TTP & \no & \no & \no & \ok & \ok\\
Force/Automatic login CSRF \cite{webspiCSF,sun2012} \cmmnt{(V.C\cite{webspiCSF}, A5\cite{sun2012})} 
& UA$\leftrightarrow$RP & \ok & \ok & \ok & \no & \no\\
IdP Mix-Up attack \cite{fett2016} \cmmnt{(3.2\cite{fett2016})}~(HTTP variant) 
&  UA$\leftrightarrow$RP & \no & \no & \ok & \no & \no\\
IdP Mix-Up attack \cite{fett2016}\cmmnt{(3.2\cite{fett2016})}~(HTTPS variant)
& UA$\leftrightarrow$RP & \ok & \indirect & \indirect & \no & \no\\
Naive session integrity attack \cite{fett2016} \cmmnt{(3.4\cite{fett2016})}
& UA$\leftrightarrow$RP & \ok & \ok & \ok & \no & \no\\
Open Redirector in OAuth 2.0 \cite{RFC6819,RFC6749} \cmmnt{(4.2.4\cite{RFC6819}, 10.15\cite{RFC6749})} 
& UA$\leftrightarrow$\{RP,TTP\} & \indirect & \indirect & \indirect & \indirect & \indirect\\
Resource Theft by Code/Token Redirection \cite{webspiCSF,WPSE} \cmmnt{(V.D\cite{webspiCSF})} 
& UA$\leftrightarrow$TTP & \ok & \no & \no & \no \cmmnt{\alert{\ok}} & \ok\\
Session swapping \cite{sun2012,RFC6749}\cmmnt{(A4\cite{sun2012}, 10.12\cite{RFC6749})} 
& UA$\leftrightarrow$RP & \ok & \ok & \ok & \no & \no\\
Social login CSRF on stateless clients \cite{webspiCSF,RFC6749}\cmmnt{(V.C\cite{webspiCSF}, 10.12\cite{RFC6749})} 
& UA$\leftrightarrow$RP & \ok & \ok & \ok & \no & \no\\
Social login CSRF through \cmmnt{AS} TTP Login CSRF \cite{webspiCSF} \cmmnt{(V.C\cite{webspiCSF})}
& UA$\leftrightarrow$TTP & \indirect & \no & \no & \ok & \ok\\
Token replay implicit mode \cite{BLAST,Wang2013,RFC6749}\cmmnt{(\cite{BLAST,Wang2013} 10.16\cite{RFC6749})}
& UA$\leftrightarrow$RP & \ok & \ok & \no & \no & \no\\
Unauth. Login by Code Redirection \cite{webspiCSF,RFC6749}\cmmnt{(V.D\cite{webspiCSF}, 10.6\cite{RFC6749})}
& UA$\leftrightarrow$TTP & \ok & \no & \no & \no & \ok\\

\hline
\multicolumn{7}{c}{\textit{PayPal}} \\
\hline

\emph{NopCommerce} gross change in IPN callback \cite{wang2011}
& RP$\leftrightarrow$\{UA,TTP\} & \no & \no & \ok & \no & \no\\
\emph{NopCommerce} gross change in PDT flow \cite{wang2011} \cmmnt{(III.1\cite{wang2011})}
& RP$\leftrightarrow$\{UA,TTP\} & \no & \no & \ok & \no & \no\\
\cmmnt{SP\(_{\text{M}}\) \emph{PayeeId} replay in SP\(_{\text{T}}\) }
Shop for free by malicious \emph{PayeeId} replay
\cite{BLAST,Pellegrino2014} \cmmnt{(\#3\cite{BLAST}, IV.A\cite{Pellegrino2014})}
& RP$\leftrightarrow$\{UA,TTP\} & \no & \no & \ok & \no & \no\\
\cmmnt{T\(_{\text{1}}\) at SP\(_{\text{T}}\) \emph{Token} replay in T\(_{\text{2}}\) at SP\(_{\text{T}}\)}
Shop for less by \emph{Token} replay
\cite{BLAST,Pellegrino2014} \cmmnt{(\#5\cite{BLAST}, IV.A\cite{Pellegrino2014})}
& UA$\leftrightarrow$RP & \no & \no & \ok & \no & \no\\

\end{tabular}
\end{small}}
\end{center}
\caption{Attacks on OAuth 2.0 and PayPal}
\label{tab:attacks}
\end{table}

In general we can see that, in the OAuth 2.0 setting, a browser extension is the most powerful tool, as it can already detect and block by itself most of the attacks (15 out of 17). The exceptions are the Facebook implicit AppId Spoofing attack~\cite{wang2012,BLAST}, which can only be detected at the TTP, and the HTTP variant of the IdP Mix-Up attack~\cite{fett2016}, which is a network attack not observable at the client. Yet, remarkably, a comparable amount of protection can be achieved by using just service workers alone: in particular, the use of service workers at both the RP and the TTP can stop 13 out of 17 attacks. The only notable differences over the browser extension approach are that: $(i)$ the code/token theft via XSS cannot be prevented, though we already discussed that even browser extensions can only partially mitigate the dangers of XSS, and $(ii)$ the \alert{resource theft by code/token redirection and the} unauthorized login by auth. code redirection cannot be stopped, because \alert{they involve} a cross-origin redirect that service workers cannot observe by SOP. Remarkably, the combination of service workers and server-side proxies offers transparent protection that goes beyond browser extensions alone: 16 out of 17 attacks are blocked, with the only exception of code/token theft via XSS as explained.

The PayPal setting shows a very different trend with respect to OAuth 2.0. Although it is possible to detect the attacks on the client side, it is not safe to do so because both browser extensions and service workers can be uninstalled by malicious customers. For example, such client-side approaches cannot prevent the shop for free attack of \cite{Pellegrino2014}, where a malicious user replaces the merchant id with her own account id. Moreover, it is worth noticing that PayPal deliberately makes heavy use of back channels (RP $\leftrightarrow$ TTP), since messages which are not relayed by the browser cannot be tampered with by malicious customers. This means that server-side proxies are the way to go to protect PayPal-like payment systems, as confirmed by the table.

\subsection{Take-Away Messages}
\label{sec:deployments}
Here we highlight the main take-away messages of our design space analysis. 
In general, we claim that different web protocols require different protection mechanisms, hence every defensive solution which is bound to a specific placement of monitors does not generalize. More specifically: 
\begin{itemize}
    \item A clear total order emerges on the ease of deployment and usability axes. Service workers score best there, closely followed by server-side proxies, whose deployment is still feasible and transparent to end-users. Browser extensions are much more problematic, especially for large-scale security enforcement. 
	\item With respect to the protection and compatibility axes, browser extensions are indeed a powerful tool, yet they can be replaced by a combination of service workers and server-side proxies to enforce transparent protection, extended to attacks which are not visible at the client alone.  
\end{itemize}

In the end, we argue that a combination of service workers and server-side proxies has the potential to reconcile security, compatibility, ease of deployment and usability. In our approach, described in the next section, we thus pursue this research direction.

\section{Proposed Approach: Bulwark}\label{sec:approach}
In this section we present Bulwark,\footnote{Bulwark is currently proprietary software at SAP: the tool could be made available upon request and an open-source license is under consideration.} our formally verified approach to the holistic security monitoring of web protocols. For space reasons, we present an informal overview and we refer to %
\alert{the online technical report for additional details~\cite{bulwarkreport}.}

\subsection{Overview}\label{sec:approach-overview}
Bulwark builds on top of ProVerif, a state-of-the-art protocol verification tool~\cite{proverif}. ProVerif was originally designed for traditional cryptographic protocols, not for web protocols, but previous work showed how it can be extended to the web setting by using the WebSpi library~\cite{webspiCSF}. In particular, WebSpi provides a ProVerif model of a standard web browser and includes support for important threats from the web security literature, e.g., \emph{web attackers}, who lack traditional Dolev-Yao network capabilities and attack the protocol through a malicious website.

Bulwark starts from a ProVerif model of the web protocol to protect, called \emph{ideal specification}, and generates formally verified security monitors deployed as service workers or server-side proxies. This process builds on an intermediate step called \emph{monitored specification}. The workflow is summarized in \figurename~\ref{fig:flow}.

\begin{figure}[t]
\centering
\resizebox{0.9\textwidth}{!}{  
  \includegraphics{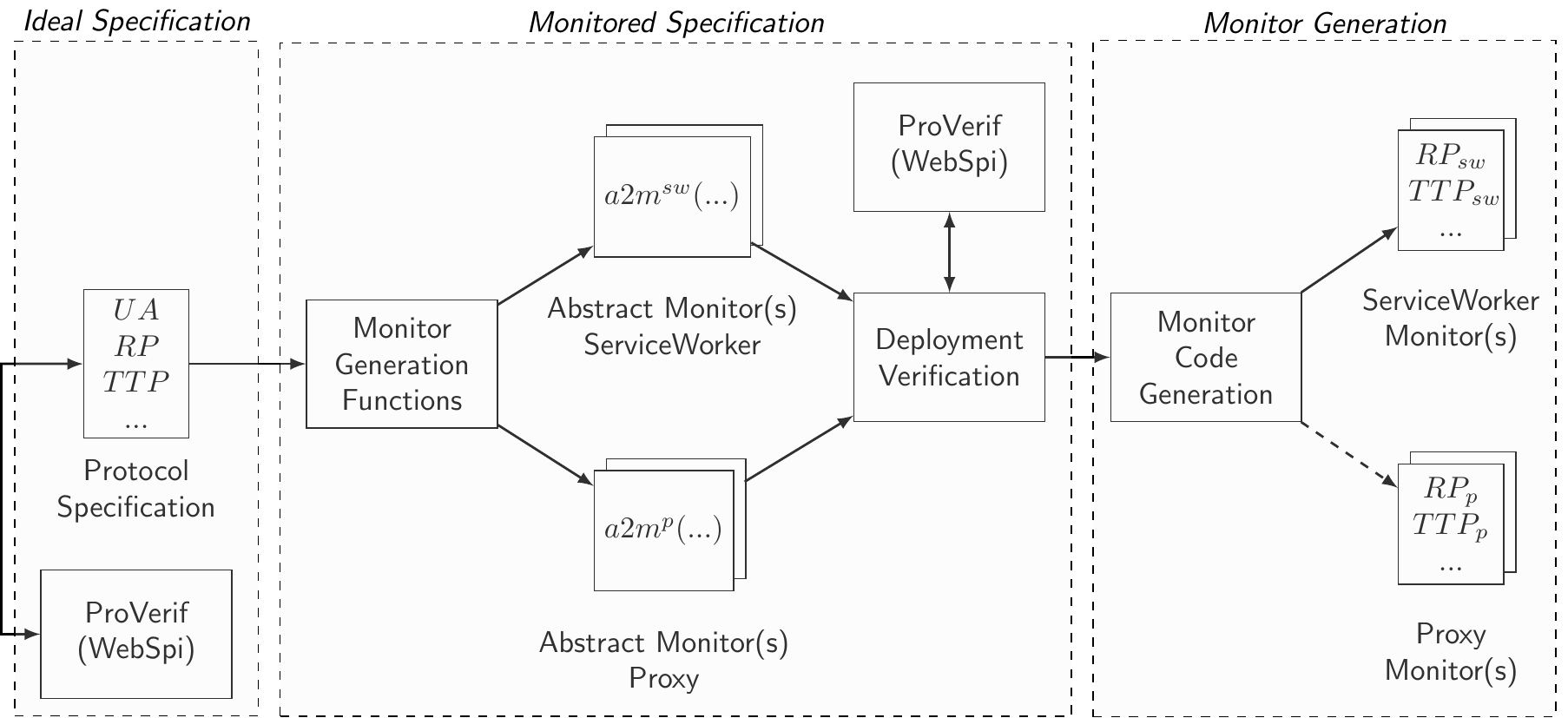}}
\caption{Monitor generation pipeline}
\label{fig:flow}
\end{figure}

To explain the intended use case of Bulwark, we focus on the typical setting of a multi-party web application including a TTP which offers integration to a set of RPs, yet the approach is general. The TTP starts by writing down its protocol in ProVerif, expressing the intended security properties by means of standard \emph{correspondence assertions} (authentication) and \emph{\alert{(syntactic)} secrecy queries} supported by ProVerif. For example, the code/token redirection attack against OAuth 2.0 (cf.~Section~\ref{sec:atk2}) can be discovered through the violation of a correspondence assertion~\cite{webspiCSF}. The protocol can then be automatically verified for security violations and the TTP can apply protocol fixes until ProVerif does not report any flaw. 
Since ProVerif is a sound verification tool~\cite{Blanchet2009}, this process eventually leads to a security proof \alert{for an unbounded number of protocol sessions}, up to the web model of WebSpi. %
\alert{The WebSpi model, although expressive, is not a complete model of the Web~\cite{webspiCSF}. For example, it does not model advanced security headers such as \texttt{Content-Security-Policy}, frames and frame communication (\texttt{postMessage}). 
However, the library models enough components of the modern Web to be able to capture all the attacks of Table~\ref{tab:attacks}}. %

Once verification is done, the TTP can use Bulwark to automatically generate security monitors for its RPs from the ideal specification, e.g., to complement the traditional protocol SDK that the TTP normally offers anyway with protection for RPs, which are widely known to be the buggiest participants~\cite{integuard}. The TTP could also decide to use Bulwark to generate its own security monitors, so as to be protected even in the case of bugs in its own implementation.

\subsection{Monitored Specification}\label{sec:approach-monitored-spec}
In the \emph{monitored specification} phase, Bulwark relaxes the ideal assumption that all protocol participants are implemented correctly. In particular, user-selected protocol participants are replaced by \emph{inattentive} variants which comply with the protocol flow, but forget relevant security checks. Technically, this is done by replacing the ProVerif processes of the participants with new processes generated by  
\alert{automatically removing from the honest participants all the security checks (pattern matches, get/insert and conditionals) on the received messages, which include the invariants represented by the boxed checks in \figurename~\ref{fig:oauth-example}.} %
This approximates the possible mistakes made by \emph{honest-but-buggy} participants, obtaining processes that are interoperable with the other participants, but whose possible requests and responses are a superset of the original ones. 
\alert{Intuitively, an inattentive participant may be willing to install a monitor to prevent attackers from exploiting the lack of forgotten security checks. On the other hand, a deliberately malicious participant has no interest in doing so.}

Then, Bulwark extracts from the ideal specification all the security invariant checks forgotten by the inattentive variants of the protocol participants and centralizes them within security monitors. This is done by applying two functions $a2m^{p}$ and $a2m^{sw}$, which derive from the participant specifications new ProVerif processes encoding security monitors deployed as a server-side proxy or a service worker respectively. The $a2m^{p}$ function is a modified version of the $a2m$ function of \cite{pironti2010}, which generates security monitors for %
cryptographic protocols.
\alert{The proxy interposes and relays messages from / to the monitored inattentive participant, after performing the intended security checks. A subtle point here is that the monitor needs to keep track of the values that are already in its knowledge and those which are generated by the monitored participant and become known only after receiving them. A security check can only be executed when all the required values are part of the monitor knowledge.}
The $a2m^{sw}$ function, instead, is defined on top of $a2m^{p}$ and the ideal $UA$ process. This recalls that a service worker is a client-side defense that acts as a reverse proxy: a subset of the checks of both the server and the client side can be encoded into the final process running on the client. \alert{The function has three main responsibilities:
\begin{enumerate*}[label=$(\roman*)$]
    \item rewriting the proxy to be compatible with the service worker API;
    \item removing the channels and values that a service worker is not able to observe;
    \item plugging the security checks made by the ideal $UA$ process into the service worker.
\end{enumerate*}
}

\begin{example}\label{ex:invariants}
\figurename~\ref{fig:invariant-example} illustrates how the $RP$ of our OAuth 2.0 example from Section~\ref{sec:example} is replaced by $I(RP)$, an inattentive variant in which the \param{state} parameter invariant is not checked. The right-hand side of the figure presents the $RP$ as in \figurename~\ref{fig:oauth-example} with a few more details on its internals, according to the ideal specification: $(i)$ upon reception of message 1, the $RP$ issues a new value for \param{state} and saves it together with the $UA$ session cookie identifier (i.e., \param{state} is bound to the client session \param{sid(UA)}); and $(ii)$ upon reception of message 7, the $RP$ checks the \param{state} parameter and its binding to the client session. The inattentive version $I(RP)$, as generated by Bulwark, is shown on the left-hand side of \figurename~\ref{fig:invariant-example}: the \param{state} is neither saved by $I(RP)$ nor checked afterward.
The left-hand side of the figure also shows the proxy $M(RP)$ generated by Bulwark as $a2m^{p}(RP)$ to enforce the \param{state} parameter invariant at $RP$. We can see that the saving and the checking of \param{state} are performed by $M(RP)$. It is worth noticing that the $M(RP)$ can only save \param{state} upon reception of message 2 from $I(RP)$. The service worker monitor $a2m^{sw}(RP)$ would look very similar to $M(RP)$. 
\end{example}

\begin{figure}[!t]
\centering
\resizebox{\textwidth}{!}{
\def\hd{0.65cm}%
\def\sp{0.1cm}%
\newcommand*{\drawframe}[3]{
  \path [frame] ($(#1)-(0.1cm,#2*\hd+2*\sp)$) -- ($(#1)-(-0.1,#2*\hd+2*\sp)$) -- ($(#1)-(-0.1cm,#3*\hd-2*\sp)$) -- ($(#1)-(0.1cm,#3*\hd-2*\sp)$) -- cycle;
}%
\begin{tikzpicture}[
    node distance=3cm and 1cm,
    font=\sffamily\footnotesize,
    box/.style={rectangle, draw ,text width = 1.2cm, fill=white, drop shadow,
      align=center, minimum height = 0.5cm},
    action/.style={rectangle,text width = 3cm, fill=gray!10, %
      align=left, minimum height = 0.5cm},
    mess/.style={font=\sffamily\scriptsize},
    frame/.style={fill=gray!10, draw=black!40}
    ]
    \node [box] (ua) {$UA$};
    \node [box, right of = ua] (mon) {$M(RP)$};
    \node [box, right of = mon] (rp) {$I(RP)$};
    \node [box, right of = rp] (ttp) {$TTP$};
    \node [box, right of = ttp, right = 1.5cm] (irp) {$RP$};
    \path[draw, dotted] (ua) -- ($(ua)-(0,11*\hd)$);
    \path[draw, dotted] (mon) -- ($(mon)-(0,11*\hd)$);
    \path[draw, dotted] (rp) -- ($(rp)-(0,11*\hd)$);
    \path[draw, dotted] (ttp) -- ($(ttp)-(0,11*\hd)$);
    \path[draw, dotted] (irp) -- ($(irp)-(0,11*\hd)$);
    \path[draw, dashed] ($(ttp)-(-1.15cm,-0.5cm)$) -- ($(ttp)-(-1.15cm,11*\hd+0.5cm)$);
    \node [action] (state1) at ($(rp)-(-1.70cm,2.5*\hd)$) {\textbf{new} state}; %
    \node [action] (state11) at ($(irp)-(-1.70cm,2.5*\hd)$) {\textbf{new} state};
    \node [action] (state2) at ($(mon)-(-1.70cm,5*\hd)$) {\textbf{save} state, sid(UA)};
    \node [action] (state22) at ($(irp)-(-1.70cm,4*\hd)$) {\textbf{save} state, sid(UA)};
    \node [action] (state3) at ($(mon)-(-1.70cm,10*\hd)$) {\textbf{check} state, sid(UA)};
    \node [action] (state33) at ($(irp)-(-1.70cm,10*\hd)$) {\textbf{check} state, sid(UA)};
    \drawframe{irp}{1.5}{6.5}
    \drawframe{irp}{9}{11.5}
    \drawframe{rp}{1.5}{4}
    \drawframe{mon}{1.5}{6.5}
    \drawframe{mon}{9}{11.5}
    \node [font=\sffamily\small] at ($(mon)-(-1.5cm,7.3*\hd)$) {...};
    \node [font=\sffamily\small] at ($(irp)-(-1.5cm,7.3*\hd)$) {...};
    \node [font=\sffamily\small] at ($(mon)-(-1.5cm,10.8*\hd)$) {...};
    \node [font=\sffamily\small] at ($(irp)-(-1.5cm,10.8*\hd)$) {...};
    \path[draw, -latex] ($(ua)-(-\sp,1.5*\hd)$) to node[above]{\circlednumber{1} Visit Login Page} ($(mon)-(\sp,1.5*\hd)$);
    \path[draw, -latex] ($(mon)-(-\sp,1.5*\hd)$) to node[above]{} ($(rp)-(\sp,1.5*\hd)$);
    \path[draw, -latex] ($(irp)-(3.5cm,1.5*\hd)$) to node[above]{\circlednumber{1} Visit Login Page} ($(irp)-(\sp,1.5*\hd)$);
    \path[draw, latex-] ($(mon)-(-\sp,4*\hd)$) to node[above]{Login Button with client\_id, reduri, \fbox{state}} ($(rp)-(\sp,4*\hd)$);
    \path[draw, latex-] ($(ua)-(-\sp,6.5*\hd)$) to node[above]{\circlednumber{2} Login Button with client\_id, reduri, \fbox{state}} ($(mon)-(\sp,6.5*\hd)$);
    \path[draw, latex-] ($(irp)-(3.5cm,6.5*\hd)$) to node[above]{\circlednumber{2} Login Button ... \fbox{state}} ($(irp)-(\sp,6.5*\hd)$);
    \path[draw, -] ($(ua)-(0,8*\hd)$) to node[above]{} ($(ttp)-(\sp,8*\hd)$);
    \path[draw, -] ($(ua)-(0,8*\hd)$) to node[above]{} ($(ua)-(0,9*\hd)$);
    \path[draw, -latex] ($(ua)-(0,9*\hd)$) to node[above]{\circlednumber{7} code, \fbox{state}} ($(mon)-(\sp,9*\hd)$);
    \path[draw, -latex] ($(irp)-(3.5cm,9*\hd)$) to node[above]{\circlednumber{7} code, \fbox{state}} ($(irp)-(\sp,9*\hd)$);
\end{tikzpicture}}

\caption{Monitor invariant example}
\label{fig:invariant-example}
\end{figure}
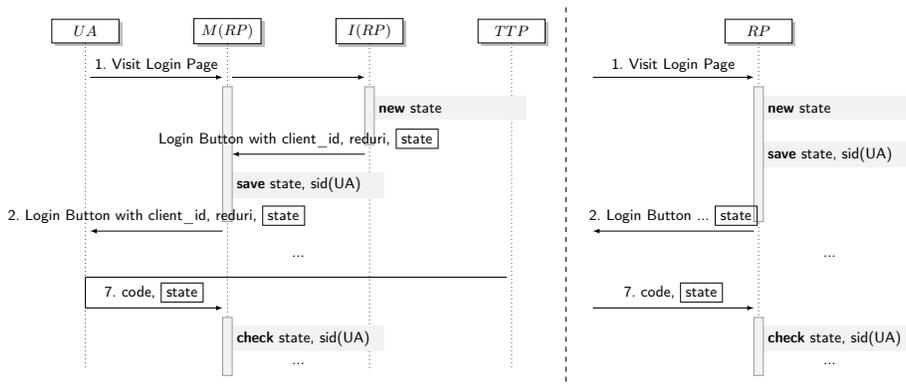

Finally, Bulwark produces a \emph{monitored specification} where each inattentive protocol participant deploys a security monitor both at the client side (service worker) and at the server side (proxy). However, this might be overly conservative, e.g., a single service worker might already suffice for security. To optimize ease of deployment, Bulwark runs again ProVerif on the possible monitor deployment options, starting from the most convenient one, until it finds a setting which satisfies the security properties of the ideal specification. As an example, consider the system in which the only inattentive participant is the $RP$. There are three possible options, in decreasing order of ease of deployment:
\begin{enumerate}
\item $TTP \parallel I(RP) \parallel ( a2m^{sw}(RP,UA) \parallel UA )$, where the monitor is deployed as a service worker registered by the $RP$ at the $UA$;
\item $TTP \parallel (I(RP) \parallel a2m^{p}(RP)) \parallel UA$, where the monitor is a proxy at $RP$;
\item $TTP \parallel  (I(RP) \parallel a2m^{p}(RP)) \parallel ( a2m^{sw}(RP,UA)
\parallel UA )$, with both.
\end{enumerate}
The first option which passes the ProVerif verification is chosen by Bulwark.

\subsection{Monitor Generation}\label{sec:approach-monitored-gen}
Finally, Bulwark translates the ProVerif monitor processes into real service workers (written in JavaScript) or proxies (written in Python), depending on their placement in the monitored specification. This is a relatively direct one-to-one translation, whose key challenge is mapping the ProVerif messages to the real HTTP messages exchanged in the web protocol. Specifically, different RPs integrating the same TTP will host the protocol at different URLs and each TTP might use different names for the same HTTP parameters or rely on different message encodings (JSON, XML, etc.).

Bulwark deals with this problem by means of a configuration file, which drives the monitor generation process by defining the concrete values of the symbols and data constructors that are used by the ProVerif model. When the generated monitor needs to apply e.g., a data destructor on a name, it searches the configuration file for its definition and calls the corresponding function that deconstructs the object into its components. 
Since data constructors/destructors are directly written in the target language as part of the monitor configuration, different implementations can be generated for the same monitor, so that a single monitor specification created for a real-world participant e.g., the Google TTP, can be easily ported to others, e.g., the Facebook TTP, just by tuning their configuration files.
%

%
%
%
%
%
%
%
%
%
%
%
%
%

%

%
%
%
%
%
%
%
%
%
%
%
%
%
%

%
%
%
%
%
%

%
%
%
%
%
%
%
%
%
%
%
%
%
%
%

%

%
%
%
%
%
%
%
%
%
%
%
%
%
%
%
%
%
%
%
%
%
%
%
%
%
%
%
%
%
%
%
%
%
%
%
%
%
%
%

%

%
%
%
%
%
%
%
%
%
%
%
%
%
%
%
%
%
%
%
%
%

%

%

%
%
%
%
%
%
%
%
%
%
%
%
%
%
%
%
%
%
%
%
%
%
%
%
%
%
%
%
%
%
%
%
%
%
%
%
%
%
%
%
%
%
%
%
%
%
%
%
%
%
%
%
%
%
%
%
%
%
%

%

%

%
%
%
%
%

%

%
%
%
%
%
%
%
%
%
%
%
%
%
%
%
%
%
%
%
%
%
%
%
%
%
%
%
%
%
%
%
%
%
%
%
%
%
%
%
%
%
%
%
%
%
%
%
%
%
%
%
%

%
%
%
%
%
%
%
%
%
%
%
%
%
%
%
%
%
%

%
%
%
%
%
%
%
%
%
%
%
%
%
%
%
%
%
%
%
%
%
%
%
%
%
%
%
%
%
%
%
%
%
%
%
%
%
%
%
%
%
%
%
%
%
%
%
%
%
%
%
%
%
%
%
%
%
%
%
%
%
%
%
%
%
%
%
%
%
%
%

%

%

%
%
%
%
%
%
%
%
%
%
%
%
%
%
%
%
%
%
%
%
%
%
%
%
%
%
%
%
%

%
%
%
%
%
%
%
%
%
%
%
%
%
%
%
%
%
%
%
%
%
%
%
%
%
%
%
%
%
%
%
%
%
%
%
%
%
%
%
%
%
%
%
%
%
%
%
%
%
%
%
%
%
%
%
%
%
%
%
%
%
%
%
%
%
%
%
%
%
%
%
%
%
%
%
%
%
%
%
%
%
%
%
%
%
%
%
%
%
%
%
%
%
%
%
%
%
%
%
%
%
%
%
%
%
%
%
%
%
%
%
%
%
%


\section{Experimental Evaluation}\label{sec:evaluation}

\subsection{Methodology}
To show Bulwark at work, we focus on the core MPWA scenarios discussed in Section~\ref{sec:attacks}. We first write ideal specifications of the OAuth 2.0 explicit protocol and the PayPal payment system in ProVerif + WebSpi. We also define appropriate correspondence assertions and secrecy queries which rule out all the attacks in Table~\ref{tab:attacks} and we apply known fixes until ProVerif is able to prove security for the ideal specifications. Then, we setup a set of case studies representative of the key vulnerabilities plaguing these scenarios (see Table~\ref{tab:test-set}). In particular, we selected vulnerabilities from Table~\ref{tab:attacks} so as to evaluate Bulwark on both the RP and TTP via a combination of proxy and service worker monitors. For each case study, we choose a set of inattentive participants and we collect network traces to define the Bulwark configuration files mapping ProVerif messages to actual protocol messages. Finally, we use Bulwark to generate appropriate security monitors and deploy them in our case studies. All our vulnerable case studies, their ideal specifications, and the executable monitors generated by Bulwark are provided as an open-source package to the community~\cite{bulwarkpackage}.

\subsubsection{Case Studies.}
We consider a range of possibilities for OAuth 2.0. We start from an entirely artificial case study, where we develop both the RP and the TTP, introducing known vulnerabilities in both parties (CS1). We then consider integration scenarios with three major TTPs, i.e., Facebook, VK and Google, where we develop our own vulnerable RPs on top of public SDKs (CS2-CS4). Finally, we consider a case study where we have no control of any party, i.e., the integration between Overleaf and Google (CS5). We specifically choose this scenario, since the lack of the state parameter in the Overleaf implementation of OAuth 2.0 introduces known vulnerabilities.\footnote{We responsibly disclosed the issue to Overleaf and they fixed it before publication.}
To evaluate the CaaS scenario, we select legacy versions of three popular e-commerce platforms, suffering from known vulnerabilities in their integration with PayPal, in particular osCommerce 2.3.1 (CS6), NopCommerce 1.6 (CS7) and TomatoCart 1.1.15 (CS8). %

\subsubsection{Evaluation criteria.}
We evaluate each case study in terms of four key aspects: $(i)$ \emph{security}: we experimentally confirm that the monitors stop the exploitation of the vulnerabilities; $(ii)$ \emph{compatibility}: we experimentally verify that the monitors do not break legitimate protocol runs; $(iii)$ \emph{portability}: we assess whether our ideal specifications can be used without significant changes across different case studies; and $(iv)$ \emph{performance}: we show that the time spent to verify the protocol and generate the monitors is acceptable for practical use.

\begin{table}[t!]
    \begin{center}
    \resizebox{\textwidth}{!}{
    \begin{tabular}{@{\makebox[2em][c]{\rownumber\space}}|c|c|c|c}
       \multicolumn{1}{@{\makebox[2em][c]{\textbf{CS}}} |c| }{\textbf{RP}} & \textbf{TTP} & \textbf{Protocol}  & \textbf{Vuln. (Tab. \ref{tab:attacks})} \\
        \hline\hline
      \textit{artificial RP 1} & \textit{artificial IdP} & OAuth 2.0 explicit & \#13 \#17 \\
      \textit{artificial RP 2} & \texttt{facebook.com} & OAuth 2.0 exp. (\texttt{graph-sdk} 5.7) & \#13  \\
      \textit{artificial RP 3} & \texttt{vk.com} & OAuth 2.0 exp. (\texttt{vk-php-sdk} 5.100) & \#13  \\
      \textit{artificial RP 4} & \texttt{google.com} & OAuth 2.0 exp. (\texttt{google/apiclient} 2.4) & \#13  \\
      \texttt{overleaf.com} & \texttt{google.com} & OAuth 2.0 explicit & \#13 \#14 \\
      \hline
      \hline
      \text{osCommerce} 2.3.1 & \texttt{paypal.com} & PayPal Standard &  \#18 \#20 \\
      \text{NopCommerce} 1.6 
      & \texttt{paypal.com} & PayPal Standard & \#18 \\
      \text{TomatoCart} 1.1.15 & \texttt{paypal.com} & PayPal Standard & \#21 \\
    \end{tabular}}
    \end{center}
    \caption{Test set of vulnerable applications}
    \label{tab:test-set} 
\end{table}

\subsection{Experimental Results}

The evaluation results are summarized in Table~\ref{tab:test-times} and discussed below. In our case studies, we considered as inattentive participants all the possible sets of known-to-be vulnerable parties, leading to 10 experiments; when multiple experiments can be handled by a single run of Bulwark, their results are grouped together in the table, e.g., the experiments for CS2-CS4. Notice that for CS1 we considered three sets of inattentive participants: only TTP (vulnerability \#17); only RP (vulnerability \#13); and both RP and TTP (both vulnerabilities). Hence, we have 3 experiments for CS1, 3 experiments for CS2-CS4, 1 experiment for CS5 and 3 experiments for CS6-CS8.

\begin{table}[t]
    \begin{center}
    \resizebox{0.9\textwidth}{!}{
    \begin{tabular}{c|l|c|c|cc|cccc|c}
        \multicolumn{4}{c|}{} & \multicolumn{2}{c}{\textbf{Monitor}} & \multicolumn{4}{|c|}{\textbf{Gen. Monitors}} &\\
         & \textbf{Ideal} & \textbf{Verification} & \textbf{Inattentive} & \multicolumn{2}{c|}{\textbf{Verification}} 
         &\multicolumn{2}{c}{\textbf{RP}} & \multicolumn{2}{c|}{\textbf{TTP}} & \textbf{Prevented} \\ 
        \textbf{CS} & \textbf{Spec.} & \textbf{Time} & \textbf{Parties} & time & \#verif. 
        & \textit{sw} & \textit{proxy} & \textit{sw} & \textit{proxy} & \textbf{Vuln.} \\ 
        \hline\hline
       \multirow{3}{*}{1} & \multirow{3}{*}{~~IS1} & \multirow{3}{*}{29m}
         & TTP & 41m & 2 & \no&\no&\no&\ok & \#17\\
       &&& RP & 15m & 1 & \ok&\no&\no&\no  & \#13\\
       &&& RP,TTP & 54m & 3 & \ok&\no&\no&\ok  & \#13 \#17\\
        \hline
       2 3 4 & ~~IS1 & 27m
         & RP & 18m & 1 & \ok&\no&\no&\no & \#13\\
       \hline
       5 & ~~IS1* & 19m
         & RP & 17m & 1 & \ok&\no&\no&\no & \#13 \#14\\
      \hline
      \hline
      6 7 8 & ~~IS2 & 3m 
        & RP & 8m & 1 & \no&\ok&\no&\no & \#18 \#20 \#21\\
    \end{tabular}}
    \end{center}
    \caption{Generated monitors and run-time}
    \label{tab:test-times}
\end{table}

\subsubsection{Security and Compatibility.}
To assess security and compatibility, we created manual tests to exploit each vulnerability of our case studies and we ran them with and without the Bulwark generated monitors. In all the experiments, we confirmed that the known vulnerabilities were prevented only when the monitors were deployed (security) and that we were able to complete legitimate protocol runs successfully both with and without the monitors (compatibility). Based on Table~\ref{tab:test-times}, we observe that 5 experiments can be secured by a service worker alone, 4 experiments can be protected by a server-side proxy and only one experiment needed the deployment of two monitors. This heterogeneity confirms the need of holistic security solutions for web protocols.

\subsubsection{Portability.}
We can see that the ideal specification IS1 created for our first case study CS1 is portable to CS2-CS4 without any change. This means that different TTPs supporting the OAuth 2.0 explicit protocol like Facebook, VK and Google can use Bulwark straightaway, by just tuning the configuration file to their settings. This would allow them to protect their integration scenarios with RPs that (like ours) make use of the state parameter. This is interesting, since different TTPs typically vary on a range of subtle details, which are all accounted for correctly by the Bulwark configuration files. However, the state parameter is not mandatory in the OAuth2 standard and thus TTPs tend to allow integration also with RPs that do not issue it. Case study CS5 captures this variant of OAuth 2.0: removing the state parameter from IS1 is sufficient to create a new ideal specification IS1*, which enables Bulwark towards these scenarios as well. As to PayPal, the ideal specification IS2 is portable to all the case studies CS6-CS8. Overall, our experience indicates that once an ideal specification is created for a protocol, then it is straightforward to reuse it on other integration scenarios based on the same protocol. 

\subsubsection{Performance.} 
We report both the time spent to verify the ideal specification (Verification Time) as well as the time needed to verify the monitors (Monitor Verification). Both steps are performed offline and just once, hence the times in the table are perfectly fine for practical adoption. Verifying the ideal specification never takes more than 30 minutes, while verifying the monitors might take longer, but never more than one hour in our experiments. The time spent in the latter step depends on how many runs of ProVerif are required to reach a secure monitored specification (see the very end of Section~\ref{sec:approach-monitored-spec}). For example, the first experiment runs ProVerif twice (cf.~\#verif.) and requires 41 minutes, while the second experiment runs ProVerif just once and thus takes only 15 minutes.

\section{Conclusion}\label{sec:conclusion}
In this paper we identified shortcomings in previous work on the security monitoring of web protocols and proposed Bulwark, the first holistic and formally verified defensive solution in this research area. Bulwark combines state-of-the-art protocol verification tools (ProVerif) with modern web technologies (service workers) to reconcile formal verification with practical security. We showed that Bulwark can generate effective security monitors on different case studies based on the OAuth 2.0 protocol and the PayPal payment system.

As future work, we plan to extend Bulwark to add an additional protection layer, i.e., on client-side communication based on JavaScript and the postMessage API. This is important to support modern SDKs making heavy use of these technologies, like the latest versions of the PayPal SDKs, yet challenging given the complexity of sandboxing JavaScript code~\cite{AckerS16}. On the formal side, we would like to strengthen our definition of ``inattentive'' participant to cover additional vulnerabilities besides missing invariant checks. For example, we plan to cover participants who forget to include relevant security headers and are supported by appropriately configured monitors in this delicate task. Finally, we would like to further engineer Bulwark to make it easier to use for people who have no experience with ProVerif, e.g., by including support for a graphical notation which is compiled into ProVerif processes, similarly to the approach in~\cite{DBLP:conf/icst/CarboneCPP15}.
%
\subsubsection*{Acknowledgments.}
Lorenzo Veronese was partially supported by the European Research Council (ERC) under the European Unions Horizon 2020 research (grant agreement No 771527-BROWSEC).

\bibliographystyle{splncs04}
\bibliography{bibliography}

\appendix
\section{Background on PayPal Standard}\label{sec:background}

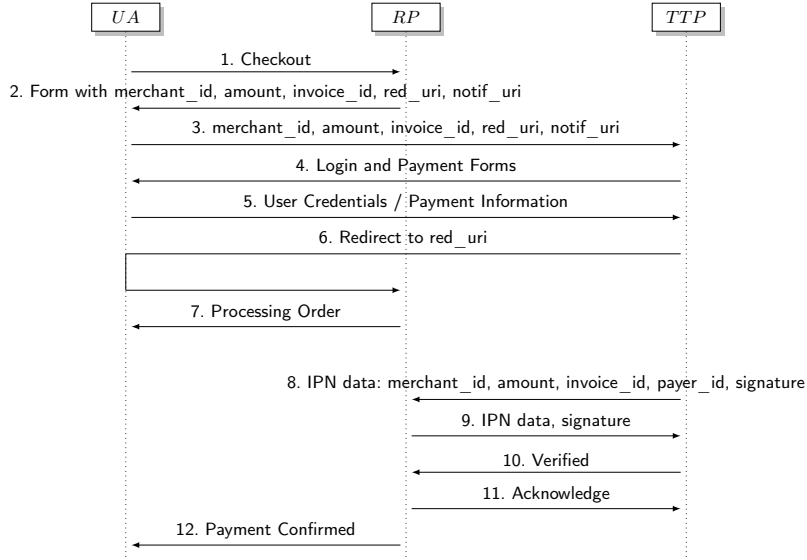
\begin{figure}[!b]
\centering
\resizebox{0.9\textwidth}{!}{
\def\hd{0.65cm}
\def\sp{0.1cm}
\begin{tikzpicture}[
    node distance=5cm and 1cm,
    font=\sffamily\footnotesize,
    box/.style={rectangle, draw ,text width = 1cm, fill=white, drop shadow,
      align=center, minimum height = 0.5cm},
    mess/.style={font=\sffamily\scriptsize}
    ]
    \node [box] (ua) {$UA$};
    \node [box, right of = ua] (rp) {$RP$};
    \node [box, right of = rp] (ttp) {$TTP$};
    \path[draw, -latex] ($(ua)-(-\sp,1.5*\hd)$) to node[above]{\circlednumber{1} Checkout} ($(rp)-(\sp,1.5*\hd)$);
    \path[draw, latex-] ($(ua)-(-\sp,2.5*\hd)$) to node[above]{\circlednumber{2} Form with merchant\_id, amount, invoice\_id, red\_uri, notif\_uri} ($(rp)-(\sp,2.5*\hd)$);
    \path[draw, -latex] ($(ua)-(-\sp,3.5*\hd)$) to node[above]{\circlednumber{3} merchant\_id, amount, invoice\_id, red\_uri, notif\_uri} ($(ttp)-(\sp,3.5*\hd)$);
    \path[draw, latex-] ($(ua)-(-\sp,4.5*\hd)$) to node[above]{\circlednumber{4} Login and Payment Forms} ($(ttp)-(\sp,4.5*\hd)$);
    \path[draw, -latex] ($(ua)-(-\sp,5.5*\hd)$) to node[above]{\circlednumber{5} User Credentials / Payment Information} ($(ttp)-(\sp,5.5*\hd)$);
    \path[draw, -] ($(ua)-(0,6.5*\hd)$) to node[above]{\circlednumber{6} Redirect to red\_uri} ($(ttp)-(\sp,6.5*\hd)$);
    \path[draw, -] ($(ua)-(0,6.5*\hd)$) to node[above]{} ($(ua)-(0,7.5*\hd)$);
    \path[draw, -latex] ($(ua)-(0,7.5*\hd)$) to node[above]{} ($(rp)-(\sp,7.5*\hd)$);
    
    \path[draw, latex-] ($(ua)-(-\sp,8.5*\hd)$) to node[above]{\circlednumber{7} Processing Order } ($(rp)-(\sp,8.5*\hd)$);
    
    \path[draw, latex-] ($(rp)-(-\sp,10.5*\hd)$) to node[above]{\circlednumber{8} IPN data: merchant\_id, amount, invoice\_id, payer\_id, signature  } ($(ttp)-(\sp,10.5*\hd)$);
    \path[draw, -latex] ($(rp)-(-\sp,11.5*\hd)$) to node[above]{\circlednumber{9} IPN data, signature  } ($(ttp)-(\sp,11.5*\hd)$);
    \path[draw, latex-] ($(rp)-(-\sp,12.5*\hd)$) to node[above]{\circlednumber{10} Verified  } ($(ttp)-(\sp,12.5*\hd)$);
    \path[draw, -latex] ($(rp)-(-\sp,13.5*\hd)$) to node[above]{\circlednumber{11} Acknowledge  } ($(ttp)-(\sp,13.5*\hd)$);
    
    \path[draw, latex-] ($(ua)-(-\sp,14.5*\hd)$) to node[above]{\circlednumber{12} Payment Confirmed } ($(rp)-(\sp,14.5*\hd)$);

    \path[draw, dotted] (ua) -- ($(ua)-(0,15*\hd)$);
    \path[draw, dotted] (rp) -- ($(rp)-(0,15*\hd)$);
    \path[draw, dotted] (ttp) -- ($(ttp)-(0,15*\hd)$);
\end{tikzpicture}}
\caption{PayPal Standard IPN Flow}
\label{fig:paypal-example}
\end{figure}

We describe here the PayPal Standard Checkout (\figurename~\ref{fig:paypal-example}), one of the most used protocols in CaaS scenarios, that distinguishes itself for the usage of Instant Payment Notification (IPN) messages on the back channel between RP and TTP. 
The protocol starts when the UA initiates the checkout at RP (step 1). 
An hidden form
is sent back to the UA, that when submitted triggers a request to the TTP (steps 2-3). The form includes: \texttt{merchant\_id}, the identifier registered for RP at TTP; \texttt{amount}, the total amount that needs to be payed; \texttt{invoice\_id}, the identifier of the current invoice; \texttt{red\_uri}, the URI at RP to which the user is redirected when the payment is completed; \texttt{notif\_uri}, the URI at RP to which the TTP will send IPN notifications. These URLs needs to be pre-registered on the PayPal website before
they are used. The UA authenticates with the TTP and confirms the payment (steps 4-5). Upon confirmation the TTP redirects the UA to the \texttt{red\_uri} of RP, that sets the status of the invoice to \textit{Processing}.
At a later point in time, the TTP sends an IPN notification to the RP \texttt{notif\_uri} to confirm the successful payment (step 8). This request contains all the previous payment information  received at step 3 and two new fields: \texttt{payer\_id}, the identifier of the user that confirmed the payment; and \texttt{signature}, a signature or MAC issued by the TTP that guarantees the message authenticity. The RP verifies the validity of the payment data in a back channel exchange with the TTP (steps 8-9) and acknowledges the reception of the notification (step 11). The RP sets the status of the invoice to \textit{Payed} and notifies the user.  Also for this protocol various attacks have been reported, e.g.,~\cite{wang2011,Pellegrino2014,BLAST}, here an example. 
\subsubsection{Shop for free by malicious \emph{PayeeId} replay \cite{BLAST,Pellegrino2014}.}
An attacker can replace the RP's PayPal account (\texttt{merchant\_id}) with her own in the first connection to the TTP at step 2.
This way the attacker can buy products at RP for free, paying herself rather than the RP. 
The RP can prevent this attack by checking that the \texttt{merchant\_id} value at step 8 matches the one that was generated at step 2.

\section{Monitored Specification and Monitor Generation}

This section provides more details about Bulwark, with special focus on the two monitor generation functions. In particular, we use again Example \ref{ex:invariants} to describe fragments of ideal specifications, inattentive participants and monitored specifications in ProVerif+WebSpi. For the complete specifications, configuration files and generated monitors we refer the interested reader to our package \cite{bulwarkpackage}.

\begin{figure}[h]
\footnotesize
\resizebox{\textwidth}{!}{
\begin{minipage}{\textwidth}
\begin{tabbing}
\phantom{0}1~ $\kwl{let}\ \var{RPApp}(\var{h}{:}\var{Host}, \var{fb}{:}\var{Host}) =  $\\
\phantom{0}2~ $\ \kwl{let}\ \var{reduri} = \var{uri}(\var{https}(), \var{h}, \kwf{callbackpath}(), \var{nullParams}())\ \kwl{in} $\\
\phantom{0}3~ $\ ( $\\
\phantom{0}4~ $\ \ \ \ (\kwl{in}(\var{httpServerRequest}, (\var{u}{:}\var{Uri}, \var{hs}{:}\var{Headers},  = \var{httpGet}(), \var{corr}{:}\kwt{bitstring})); $\\
\phantom{0}5~ $\ \ \ \ \ \kwl{let}\ \var{uri}( = \var{https}(),  = \var{h},  = \kwf{loginpath}(),  = \var{nullParams}()) = \var{u}\ \kwl{in} $\\
\phantom{0}6~ $\ \ \ \ \ \kwl{let}\ \var{cp} = \kwf{session{\_}start}(\var{getCookie}(\var{hs}), \var{corr})\ \kwl{in} $\\
\phantom{0}7~ $\ \ \ \ \ \kwl{new}\ \var{state}{:}\kwt{bitstring}; $\\
\phantom{0}8~ $\ \ \ \ \ \kwl{insert}\ \kwt{RPSessions}(\var{cp}, \var{state}); $\\
\phantom{0}9~ $\ \ \ \ \ \kwl{let}\ \var{fb{\_}uri} = \var{uri}(\var{https}(), \var{fb}, \kwf{oauthpath}(), \kwf{codereqparams}(\kwc{appid}, \var{reduri}, \var{state}))\ \kwl{in} $\\
10~ $\ \ \ \ \ \kwl{event}\ \kwe{rp{\_}begin}(\var{h}, \var{fb}, \var{cp}, \kwc{appid}, \var{reduri}, \var{state}); $\\
11~ $\ \ \ \ \ \kwl{out}(\var{httpServerResponse}, (\var{u}, \var{httpOk}(\kwf{pagewithlink}(\var{fb{\_}uri})), \var{cp}, \var{unsafeUrl}(), \var{corr}))) $\\
12~ $\mid $\\
13~ $\ \ \ \ (\kwl{in}(\var{httpServerRequest}, (\var{u}{:}\var{Uri}, \var{hs}{:}\var{Headers},  = \var{httpGet}(), \var{corr}{:}\kwt{bitstring})); $\\
14~ $\ \ \ \ \ \kwl{let}\ \var{uri}( = \var{https}(),  = \var{h},  = \kwf{callbackpath}(), \kwf{coderesparams}(\var{code}, \var{state})) = \var{u}\ \kwl{in} $\\
15~ $\ \ \ \ \ \kwl{get}\ \kwt{RPSessions}( = \var{getCookie}(\var{hs}), = \var{state})\ \kwl{in} $\\
16~ $\ \ \ \ \ \kwl{let}\ \var{req{\_}uri} = \var{uri}(\var{https}(), \var{fb}, \kwf{tokenpath}(), \kwf{tokenreqparams}(\kwc{appid}, \var{reduri}, \kwc{appsecret}, \var{code}))\ \kwl{in} $\\
17~ $\ \ \ \ \ \kwl{new}\ \var{ncorr}{:}\kwt{bitstring}; $\\
18~ $\ \ \ \ \ \kwl{out}(\var{httpServerRequest}, (\var{req{\_}uri}, \var{headers}(\var{noneUri}, \var{nullCookiePair}(), \var{notajax}()), \var{httpGet}(), \var{ncorr})); $\\
19~ $\ \ \ \ \ \kwl{in}(\var{httpServerResponse}, ( $\\
\phantom{20~} $\ \ \ \ \ \ \ \ = \var{req{\_}uri},  \var{httpOk}(\kwf{tokenresjson}(\var{token})), \var{cp1}{:}\var{CookiePair},  \var{rp1}{:}\var{ReferrerPolicy},  = \var{ncorr})); $\\
20~ $\ \ \ \ \ \kwl{event}\ \kwe{rp{\_}end}(\var{h}, \var{fb}, \var{cp}, \kwc{appid}, \var{reduri}, \kwc{appsecret}, \var{state}, \var{code}, \var{token}); $\\
21~ $\ \ \ \ \ \kwl{out}(\var{httpServerResponse}, (\var{u}, \var{httpOk}(\kwf{success}()), \var{cp}, \var{noReferrer}(), \var{corr})))). $%
\end{tabbing}%
\end{minipage}}
\caption{RP Process}
\label{fig:rpprocess}
\end{figure}

\subsection{Ideal Specification.} 
We model the protocol as a set of ProVerif+WebSpi processes and queries~\cite{proverifmanual}. \figurename~\ref{fig:rpprocess} presents the model of the RP process that handles two requests: an HTTP GET to its login path (message 1 of \figurename~\ref{fig:invariant-example} $\rightarrow$ lines 4-11 of \figurename~\ref{fig:rpprocess}) and a GET to the OAuth callback path (message 7  of \figurename~\ref{fig:invariant-example} $\rightarrow$ lines 13-20 of \figurename~\ref{fig:rpprocess}). The \textbf{new}, \textbf{save} and \textbf{check} operations of \figurename~\ref{fig:invariant-example} respectively correspond to lines 7, 8 and 15 of \figurename~\ref{fig:rpprocess}. We use a ProVerif table to represent the session storage of the RP, that is indexed by the user cookie ($cp$) (the equivalent of \texttt{sid(UA)} in \figurename~\ref{fig:invariant-example}).
The \texttt{event} statements at lines 10 and 20 are part of the security specification and explicitly label security relevant events on which queries are defined.
As an example, the following query, that represents a typical authentication property, states that when the UA reaches the end of the protocol in the callback path of RP, there need to be a corresponding explicit start of the protocol at RP:
\begin{tabbing}%
\footnotesize
$\kwl{query}\ %
...; $
$\ \kwl{event}(\ \kwe{ua{\_}end}(\var{b}, \var{h}, \var{idph}, \var{state}, \var{code}) $ \\
\footnotesize
$\ \wedge\ \kwl{event}(\ \kwe{rp{\_}end}(\var{h}, \var{idph}, \var{c}, \var{aid}, \var{reduri}, \var{sec}, \var{state}, \var{code}, \var{token})\ ) $\\
\footnotesize
$\ \Longrightarrow 
\kwl{event}\ (\ \kwe{rp{\_}begin}(\var{h}, \var{idph}, \var{c}, \var{aid}, \var{reduri}, \var{state})\ )). $%
\end{tabbing}%

\subsection{Monitored Specification}

\subsubsection{Inattentive Participants.} 
The inattentive RP generated by Bulwark is obtained by removing every \texttt{insert}, \texttt{get} and pattern match from the process in \figurename~\ref{fig:rpprocess} except those that select the URL path to handle just after the reception of a message (lines 5, 14). This is the case because the inattentive RP still needs to be interoperable with the original process.

\subsubsection{Proxy Monitors.}

\begin{algorithm}[t]
\caption{Proxy Generation Function (Excerpt)}
\label{alg:a2m}
\renewcommand{\algorithmicrequire}{\textbf{Input:}}
\renewcommand{\algorithmicensure}{\textbf{Variables:}}
\begin{algorithmic}[1]
\Require Process $P$
\Ensure $known = \emptyset$, $buffers = \emptyset$, $delayedExps = \emptyset$
\Procedure{$a2m^p$}{$P$}
\If{$P$ \textbf{is} $\textbf{0}$}
    \State \textbf{return} $flushBuffers()$; \textbf{0} %
\ElsIf{$P$ \textbf{is} $\texttt{let } d = t \texttt{ in } P'$}
    \If{$t \in known$}
        \State $known \gets known \cup \{d\}$
        \State \textbf{return} $\texttt{let } d = t \texttt{ in }  a2m^{p}(P')$
    \EndIf
        \State $delayedExps \gets delayedExps \cup \{  \texttt{let } d = t \}$
        \State \textbf{return} $a2m^{p}(P')$
\ElsIf{$P$ \textbf{is} $\texttt{insert } table(t); P'$}
    \If{$t \in known$}
        \State \textbf{return} $\texttt{insert } table(t); a2m^{p}(P')$
    \EndIf
    \State $delayedExps \gets delayedExps \cup \{  \texttt{insert } table(t) \}$
    \State \textbf{return} $a2m^{p}(P')$
\ElsIf{$P$ \textbf{is} $\texttt{new } a;P'$}
    \State \textbf{return} $a2m^{p}(P')$
\ElsIf{$P$ \textbf{is} $\texttt{in}(c, d);P'$}
    \State $known \gets known \cup \{d\}$
    \State $buffers \gets buffers \cup \{ (mch(c), d) \}$
    \State \textbf{return} $\texttt{in}(c, d); a2m^{p}(P')$
\ElsIf{$P$ \textbf{is} $\texttt{out}(c, t);P'$}
    \State $known \gets known \cup \{t\}$
    \State \textbf{return} $flushBuffers(); $
    \State $\quad \texttt{in}(mch(c), t); doChecks(delayedExps); \texttt{out}(c, t); a2m^{p}(P')$
\EndIf
\EndProcedure
\end{algorithmic}
\end{algorithm}

Once Bulwark has generated the inattentive variant of the protocol, it applies the $a2m^{p}$ function (Algorithm~\ref{alg:a2m}) to the ideal RP process.
The function is a modified version of the $a2m$ function of \cite{pironti2010}. It takes as input a ProVerif process $P$ and returns the associated proxy monitor process. Specifically, each time $P$ sends / waits for data on the channel $c$, the monitor interposes and relays the message from / to $P$ over a new channel $mch(c)$, after performing appropriate security checks. The function makes use of three variables: $known$ tracks the values that are part of the knowledge of the monitor; $buffers$ tracks all the messages that are received by the monitor and needs to be relayed to the process; $delayedExps$ tracks the expressions that cannot be immediately executed by the monitor since they predicate on values that are not part of the $known$ variables. When the knowledge is updated with the correct values, the monitor applies these delayed expressions to the newly available data.

We describe the function by examples, showing how lines 4 - 11 of \figurename~\ref{fig:rpprocess} are translated by Algorithm~\ref{alg:a2m}. Part of the output process is shown in \figurename~\ref{fig:rpproxyprocess}. Note that Bulwark applies some minor optimizations to the output process, such as removing unused variables and applying some rewriting rules to destructors and constructors.
\begin{itemize}
\item \textit{Lines 4-6 of \figurename~\ref{fig:rpprocess}, Lines 4-6 of \figurename~\ref{fig:rpproxyprocess}}: The input process receives an HTTP request and does a pattern-match on the URL to select the login path. 
For every $\texttt{in}(c,d)$ that is executed by the input process, a corresponding \texttt{in} is executed by the monitor. This operation increases the knowledge of the monitor by the value $d$ and the received data is buffered (lines 18-21 of Alg.~\ref{alg:a2m}).

\item \textit{Lines 7-9 of \figurename~\ref{fig:rpprocess}}: The process generates a new \texttt{state} parameter, it inserts it into the session storage and creates a login URI.
The \texttt{new} operations are not executed by the monitor, they can only be executed by the monitored party (lines 16-17 of Alg.~\ref{alg:a2m}).
The \texttt{insert} statement at line 8 cannot be executed by the monitor since it does not know yet the value of \texttt{state} that has been generated by the monitored entity: the expression is thus delayed (lines 11-15 of Alg.~\ref{alg:a2m}). The same applies for the \texttt{let} at line 9.

\item \textit{Line 11 of \figurename~\ref{fig:rpprocess}, Lines 7-14 \figurename~\ref{fig:rpproxyprocess}}: The process sends an HTTP response containing the generated URI.
For every $\texttt{out}(c,t)$ of the input process, the monitor first sends it all the buffered data (line 7 of \figurename~\ref{fig:rpproxyprocess}) then waits for the monitored application to send $t$ on the channel between the monitor and the application ($mch(c)$).
When $t$ is received (line 8 of \figurename~\ref{fig:rpproxyprocess}), the knowledge of the monitor is increased by $t$. This enables the monitor to execute all the expressions that have been delayed in the previous steps ($doChecks(delayedExps)$).
This includes the \texttt{insert} statement at line 13 (\figurename~\ref{fig:rpproxyprocess}).
Finally the monitor executes the \texttt{out} operation (lines 22-25 of Alg.~\ref{alg:a2m}).
\end{itemize}
\begin{figure}[t]
\footnotesize
\resizebox{\textwidth}{!}{
\begin{minipage}{\textwidth}
\begin{tabbing}
\phantom{0}1~ $\kwl{let}\ \var{RPProxy}(\var{h}{:}\var{Host}, \var{fb}{:}\var{Host}) =  $\\
\phantom{0}2~ $\ \kwl{let}\ \var{reduri} = \var{uri}(\var{https}(), \var{h}, \kwf{callbackpath}(), \var{nullParams}())\ \kwl{in} $\\
\phantom{0}3~ $\ (\  $\\
\phantom{0}4~ $\ \ (\kwl{in}(\var{httpServerRequest}, (\var{u}{:}\var{Uri}, \var{hs}{:}\var{Headers}, \var{cs{\_}1001}{:}\var{HttpRequest}, \var{corr}{:}\kwt{bitstring})); $\\
\phantom{0}5~ $\ \ \ \kwl{let}\ (\var{uri}( = \var{https}(),  = \var{h},  = \var{loginpath}(),  = \var{nullParams}())) = \var{u}\ \kwl{in} $\\
\phantom{0}6~ $\ \ \ \kwl{let}\ ( = \var{httpGet}()) = \var{cs{\_}1001}\ \kwl{in} $\\
\phantom{0}7~ $\ \ \ \kwl{out}(\var{mC{\_}1{\_}out}, (\var{u}, \var{hs}, \var{cs{\_}1001}, \var{corr})); $\\
\phantom{0}8~ $\ \ \ \kwl{in}(\var{mC{\_}1{\_}in}, (\var{cs{\_}1102}{:}\var{Uri}, \var{cs{\_}1100}{:}\var{HttpResponse}, \var{cp}{:}\var{CookiePair}, \var{cs{\_}1101}{:}\var{ReferrerPolicy},  = \var{corr})); $\\
\phantom{0}9~ $\ \ \ \kwl{let}\ (\var{httpOk}(\var{pagewithlink}(\var{uri}( = \var{https}(),  = \var{fb},  = \var{oauthpath}(), $\\
\phantom{10~} $\ \ \ \ \ \ \ \ \ \ \ \ \ \ \ \ \ \ \ \ \var{codereqparams}( = \var{appid},  = %
10~ %
11~ \var{reduri}, \var{state}))))) = \var{cs{\_}1100}\ \kwl{in} $\\
12~ $\ \ \ \kwl{let}\ ( = \var{uri}(\var{https}(), \var{h}, \var{loginpath}(), \var{nullParams}())) = \var{cs{\_}1102}\ \kwl{in} $\\
13~ $\ \ \ \kwl{insert}\ \var{MRPSessions}(\var{cp}, \var{state}); $\\
14~ $\ \ \ \kwl{out}(\var{httpServerResponse}, (\var{cs{\_}1102}, \var{cs{\_}1100}, \var{cp}, \var{cs{\_}1101}, \var{corr}))) $\\
15~ $\ \mid $\\
16~ $\ \ (\kwl{in}(\var{httpServerRequest}, (\var{u}{:}\var{Uri}, \var{hs}{:}\var{Headers}, \var{cs{\_}1001}{:}\var{HttpRequest}, \var{corr}{:}\kwt{bitstring})); $\\
17~ $\ \ \ \kwl{let}\ (\var{uri}( = \var{https}(),  = \var{h},  = \var{callbackpath}(), \var{coderesparams}(\var{code}, \var{state}))) = \var{u}\ \kwl{in} $\\
18~ $\ \ \ \kwl{let}\ ( = \var{httpGet}()) = \var{cs{\_}1001}\ \kwl{in} $\\
29~ $\ \ \ \kwl{get}\ \var{MRPSessions}( = \var{getCookie}(\var{hs}),  = \var{state})\ \kwl{in} $\\
20~ $\ \ \ \kwl{out}(\var{mC{\_}1{\_}out}, (\var{u}, \var{hs}, \var{cs{\_}1001}, \var{corr})); $\\
\phantom{22~} $\ \ \ ...$%
\end{tabbing}%
\end{minipage}}
\caption{RP Proxy Process}
\label{fig:rpproxyprocess}
\end{figure}

In summary, the proxy receives HTTP connections in place of the monitored RP (lines 4 or 16), then, depending on the values that are available in the request it could decide to execute some checks (as in line 29) or to forward the request (lines 7 and 20). When the request is forwarded, it waits for the response from the monitored application, then it executes the remaining invariant checks (as in lines 9-13) and sends the response to the UA (line 14).

\subsubsection{Service Worker Monitors.}

\begin{figure}[t]
\footnotesize
\centering
\resizebox{\textwidth}{!}{
\begin{minipage}{\textwidth}
\begin{tabbing}
\phantom{0}1~ $\kwl{let}\ \var{RPServiceWorker}(b:Browser) =  $\\
\phantom{0}2~ $\ \kwl{let}\ \var{reduri} = \var{uri}(\var{https}(), \var{h}, \var{callbackpath}(), \var{nullParams}())\ \kwl{in} $\\
\phantom{0}3~ $\ \kwl{in}(\var{serviceWorkerFetch}(\var{b}), (\var{u}{:}\var{Uri}, \var{cs{\_}1000}, \var{sw{\_}ref}{:}\var{Uri}, \var{sw{\_}p}{:}\var{Page}, \var{sw{\_}aj}{:}\var{Ajax})); $\\
\phantom{0}4~ $\ \ \kwl{let}\ ( = \var{httpGet}()) = \var{cs{\_}1000}\ \kwl{in} $\\
\phantom{0}5~ $\ \ \kwl{let}\ (\var{uri}( = \var{https}(),  = \var{h},  = \var{loginpath}(),  = \var{nullParams}())) = \var{u}\ \kwl{in} $\\
\phantom{0}6~ $\ \ (\ \kwl{out}(\var{rawRequest}(\var{b}), (\var{u}, \var{cs{\_}1000}, \var{sw{\_}ref}, \var{sw{\_}p}, \var{sw{\_}aj})); $\\
\phantom{0}7~ $\ \ \ \ \kwl{in}(\var{serviceWorkerResult}(\var{b}), ( = \var{u}, \var{cs{\_}1100}{:}\var{HttpResponse}, \var{cs{\_}1101}{:}\var{ReferrerPolicy},\ \var{xd}{:}\var{XDR}, \var{corr}{:}\kwt{bitstring})); $\\
\phantom{0}8~ $\ \ \ \ \kwl{let}\ (\var{httpOk}(\var{pagewithlink}(\var{uri}( = \var{https}(),  = \var{fb},  = \var{oauthpath}(),  $\\
\phantom{11~} $\ \ \ \ \ \ \ \ \ \ \ \ \ \ \ \ \ \ \ \ \ \ \ \ \ \ \ \ \ \var{codereqparams}( = \var{appid},  = \var{reduri}, \var{state}))))) = \var{cs{\_}1100}\ \kwl{in} $\\
\phantom{0}9~ $\ \ \ \ \kwl{insert}\ \var{MRPSessions}(\var{b}, \var{state}); $\\
10~ $\ \ \ \ \kwl{out}(\var{serviceWorkerSendHttpResponse}(\var{b}), (\var{u}, \var{cs{\_}1100}, \var{cs{\_}1101}, \var{xd}, \var{corr}))) $\\
11~ $\ \kwl{else}\ \kwl{let}\ (\var{uri}( = \var{https}(),  = \var{h},  = \var{callbackpath}(), \var{coderesparams}(\var{code}, \var{state}))) = \var{u}\ \kwl{in} $\\
12~ $\ \ (\ \kwl{get}\ \var{MRPSessions}(= \var{b}, = \var{state})\ \kwl{in} $\\
13~ $\ \ \ \ \kwl{out}(\var{rawRequest}(\var{b}), (\var{u}, \var{cs{\_}1000}, \var{sw{\_}ref}, \var{sw{\_}p}, \var{sw{\_}aj})); $\\
14~ $\ \ \ \ \kwl{in}(\var{serviceWorkerResult}(\var{b}), ( = \var{u}, \var{cs{\_}1200}{:}\var{HttpResponse}, \var{cs{\_}1201}{:}\var{ReferrerPolicy}, \var{xd}{:}\var{XDR}, \var{corr}{:}\kwt{bitstring})); $\\
15~ $\ \ \ \ \kwl{let}\ (\var{httpOk}(\var{success}())) = \var{cs{\_}1200}\ \kwl{in} $\\
16~ $\ \ \ \ \kwl{out}(\var{serviceWorkerSendHttpResponse}(\var{b}), (\var{u}, \var{cs{\_}1200}, \var{noReferrer}(), \var{xd}, \var{corr}))). $%
\end{tabbing}%
\end{minipage}}
\caption{RP Service Worker Process}
\label{fig:rpmonitorprocess}
\end{figure}

Bulwark then applies the $a2m^{sw}$ function to the RP process. 
The function is defined in terms of the $a2m^{p}$ and its main responsibilities are:
\begin{enumerate*}[label=(\roman*)]
    \item rewriting the proxy to be compatible with the service worker API;
    \item removing the channels and values that a service worker is not able to observe;
    \item joining the checks made by the UA process into the service worker.
\end{enumerate*}
The output process is shown in \figurename~\ref{fig:rpmonitorprocess}. It first receives a fetch event (line 3), then branches on the URL to select which path to handle using ProVerif \texttt{let}/\texttt{else} construct (lines 5 and 11).
The service worker differs from the reverse proxy in the channels it is able to observe and in the values it have access to. Service workers do not have access to back channels and can make http request only using the fetch API ($rawRequest$/$serviceWorkerResult$) (lines 6 and 13).
Moreover, service workers do not have access to cookies, but they are implicitly bound to the browser session: a service worker can use the browser handle to model this implicit session (lines 9 and 12). %

\subsection{Code Generation}
Bulwark' generated monitors require an input configuration file to map the symbols and data constructors used in the ProVerif messages to actual protocol messages.  \figurename~\ref{fig:swconfig} shows a fragment of the configuration file used by the RP's service worker monitor.  The mapping is straightforward for most of the symbols: e.g., the abstract symbol \texttt{h} is mapped to the string \texttt{integrator.com}. For certain symbols slightly more complex mapping operation may be required: e.g., the deconstruction of the query string parameters \texttt{codereqparams} boils down to execute the trivial JavaScript function within lines 11-14. 

\begin{figure}[t]
  \begin{lstlisting}[language=java]
const h = "integrator.com"
const fb = "www.facebook.com"
const loginpath = "/login"
const callbackpath = "/fb-callback"
const oauthpath = "/v3.2/dialog/oauth"
const appid = "390639"

let db = new zango.Db('SW', { MRPSessions: ['col_1'] })
let MRPSessions = db.collection('MRPSessions')

const codereqparams = (qs) => {
  let params = parseQuery(qs)
  return [ 
    params['client_id'], new URL(params['redirect_uri']), params['state'] ] }
  \end{lstlisting}%
\caption{Service Worker Monitor Configuration file for RP (excerpt)}
\label{fig:swconfig}
\end{figure}

Let us see now the mapping at work considering the URL deconstruction operation that the RP service worker monitor needs to execute at line 8 of \figurename~\ref{fig:rpmonitorprocess}:
\begin{Verbatim}[fontsize=\small,commandchars=\\\{\}]
uri(=https(),=\tca{fb},=\tcb{oauthpath()},\tcc{codereqparams(=appid,=reduri,state)})
\end{Verbatim}
where the \texttt{$=$} symbol indicates that a pattern matching is required. 
If the following concrete URL is received
\begin{Verbatim}[fontsize=\small,commandchars=\\\{\}]
https://\tca{www.facebook.com}\tcb{/v3.2/dialog/oauth}?\tcc{client_id=390639 &} 
  \tcc{redirect_uri=integrator.com/fb-callback & state=5d938a}
\end{Verbatim}
then the monitor would deconstruct it using first the predefined built-in function  \texttt{uri()} that would extract the four elements \texttt{https}, \tca{www.facebook.com}, \tcb{/v3.2/dialog/oauth} and \tcc{client\_id=390639 \& redirect\_uri=integrator.com/fb-callback \& state=5d938a}. The first three are successfully matched with the values associated to the abstract symbols \texttt{https()} (always equal to \texttt{https}), \tca{fb} (cf. line 2 in \figurename~\ref{fig:swconfig}), and \tcb{oauthpath()} (cf. line 5 in \figurename~\ref{fig:swconfig}). The last element is further deconstructed using \tcc{\texttt{codereqparams()}} and so on and so forth.

%
%
%
%
%

%

%
%
%
%
%
%
%
%
%
%

%

%
%
%
%
%
%
%
%
%

%
%
%
%
%
%
%
%
%
%
%
%
%
%
%
%
%
%
%
%
%


\endgroup

\end{document}